\documentclass[prd,showpacs,superscriptaddress,nofootinbib,floatfix,11pt]{revtex4-2}
\usepackage[utf8]{inputenc}
\usepackage{amsmath,amssymb}
\usepackage{slashed}
\usepackage{subfigure}
\usepackage[colorlinks=true,
breaklinks=true,
urlcolor=magenta,
citecolor=blue]{hyperref}
\usepackage[usenames,dvipsnames]{color}
\usepackage{appendix}
\usepackage{braket,bm}
\usepackage{multirow}
\usepackage{enumitem}
\usepackage{color}
\usepackage{slashed}
\usepackage{orcidlink}
\usepackage{graphicx}
\usepackage{tikz}
\usepackage{booktabs}
\usepackage{array}
\usepackage{overpic}
\usepackage{float}
\usepackage{threeparttable}
\allowdisplaybreaks[4]

\newcommand{\zjwl}{\affiliation{Junior College, Zhejiang Wanli University, Zhejiang 315101, China}}

\newcommand{\nbu}{\affiliation{Physics Department, Ningbo University, Zhejiang 315211, China}}

\newcommand{\bnu}{\affiliation{School of Physics and astronomy, Beijing Normal University, Beijing 100875, China}}

\newcommand{\ictp}{\affiliation{The Abdus Salam ICTP, Stradan Costiera 11, 34151, Trieste, Italy}}

\newcommand{\usc}{\affiliation{School of Nuclear Science and Technology, University of South China, Hengyang, 421001, Hunan, China}}

\begin{document}
\title{Possible bound states of Heavy Baryonium and Heavy Dibaryon systems} 
\author{Jing-Juan Qi\orcidlink{0000-0002-9260-9408}}\email{qijj@mail.bnu.edu.cn}
\zjwl\nbu
\author{Zhen-Hua Zhang\orcidlink{0000-0001-5031-9499}}\email{zhangzh@usc.edu.cn}
\usc
\author{Xin-Heng Guo\orcidlink{0000-0002-9309-9112}}\email{xhguo@bnu.edu.cn}
\bnu\ictp
\author{Zhen-Yang Wang\orcidlink{0000-0002-4074-7892}}\email{wangzhenyang@nbu.edu.cn}
\nbu

\begin{abstract}
In this work, we systematically study the heavy baryonium and heavy dibaryon systems using the Bethe-Salpeter equation in the ladder and instantaneous approximations for the kernel. Our results indicate that all the heavy baryonium systems, specifically $\Lambda_Q\bar{\Lambda}_Q$, $\Xi_Q\bar{\Xi}_Q$, $\Sigma_Q\bar{\Sigma}_Q$, $\Xi'_Q\bar{\Xi}'_Q$, and $\Omega_Q\bar{\Omega}_Q$ ($Q=c, b$), can form bound states. Among the heavy dibaryon systems, only the $\Xi_Q\Xi_Q$ system with $I=0$ and the $\Sigma_Q\Sigma_Q$ systems with $I=0$ and $I=1$ can exist as bound states. Additionally, the $\Sigma_Q\bar{\Sigma}_Q$ system with $I=2$ and the $\Sigma_Q\Sigma_Q$ system with $I=1$ are not deeply bound.
\end{abstract}
\maketitle
\newpage

\section{Introduction}
\label{intro}
Since the discovery of the $X(3872)$ by the Belle Collaboration in 2003 \cite{Belle:2003nnu}, numerous exotic states have been discovered in the charmed sector by various experiments, including BES$\mathrm{\uppercase\expandafter{\romannumeral3}}$, BaBar, Belle, D0, ATLAS and LHCb, et al. (see, e.g., Refs. \cite{Hosaka:2016pey,Meng:2022ozq,Chen:2022asf,Brambilla:2019esw,Liu:2019zoy,Ali:2017jda,Guo:2017jvc,Olsen:2017bmm,Lebed:2016hpi,Esposito:2016noz} for recent reviews). A common feature of these exotic states is that their masses are mostly located near the threshold of two hadrons. For example, $X(3872)$ and $Z_c(3900)$ are near the $D\bar{D}^\ast$ threshold, $T^+_{cc}$ is near the $DD^\ast$ threshold, $Z_{cs}(3985)$ is near the $D^\ast\bar{D}_s$ and $D\bar{D}^\ast_s$ thresholds, $P_c$ states are near the $\bar{D}^{(\ast)}\Sigma_c$ thresholds, and $X(6900)$ is near the $\chi_{c0}\chi_{c1}$ threshold. Therefore, these exotic hadrons are naturally considered as candidates for hadronic molecular states and believed to have four or five quarks. Their exotic spectra and decay widths have made them popular and intriguing topics in both theoretical and experimental research, deepen our understanding of the nature of QCD. 

Many heavy tetraquark and pentaquark states have already been discovered. Therefore, it is urgent to extend the research to heavy hexaquark states. The existence of the baryon-antibaryon (baryonium) and the baryon-baryon (dibaryon) molecular states has naturally become a significant research topic. In the light hexaquark sector, the deuteron is a well known molecular state composed of a proton and a neutron, with a binding energy of 2.225 MeV \cite{Weinberg:1962hj,Weinberg:1963zza,Weinberg:1965zz}. Recently, the BES$\mathrm{\uppercase\expandafter{\romannumeral3}}$ experiment group reported the observation of a $p\bar{p}$ bound state in the $3(\pi^+\pi^-)$ invariant mass spectrum \cite{BESIII:2023vvr}, which has been predicted by many theoretical works to favor decays into the final states with these pions \cite{Yan:2004xs,Ding:2005ew,Yang:2022kpm}. In the charm sector, the Belle Collaboration observed $Y(4630)$ (located 61 MeV above the $\Lambda_c\bar{\Lambda}_c$ threshold) in the $e^+e^-\rightarrow\Lambda_c\bar{\Lambda}_c$ process in 2008 \cite{Belle:2008xmh}, but no resonance structure was observed around 4.63 GeV by the BES$\mathrm{\uppercase\expandafter{\romannumeral3}}$ Collaboration \cite{BESIII:2023rwv}. The nature of $Y(4630)$ as a $\Lambda_c\bar{\Lambda}_c$ molecular state is highly debated in theory \cite{Cotugno:2009ys,Guo:2010tk,Simonov:2011jc,Liu:2021gva,Agaev:2022iha,Song:2022yfr,Mei:2022msh,Salnikov:2023cug,Wang:2013exa,Liu:2016sip,Guo:2016iej,Wang:2016fhj,Dai:2017fwx,Sundu:2018toi,Anwar:2018sol,Cao:2019wwt}. Compared with light baryon molecules, the larger masses of heavy baryons reduce the system's kinetic energy, facilitating the formation of molecules. Thus, the existence of heavy baryon molecules has attracted significant theoretical interest, and has been studied through various models such as the chiral constituent quark model \cite{Carames:2015sya,Garcilazo:2020acl}, the color flux-tube model \cite{Deng:2013aca}, the quark delocalization color screening model \cite{Huang:2013rla}, lattice QCD \cite{Junnarkar:2022yak}, chiral effective field theory \cite{Chen:2022iil,Lu:2017dvm,Oka:2013xxa,Chen:2013sba,Chen:2011cta}, QCD sum rules \cite{Wang:2021pua,Wang:2021qmn,Wan:2019ake}, the one-boson-exchange model \cite{Cheng:2022vgy,Ling:2021asz,Chen:2017vai,Meguro:2011nr,Lee:2011rka,Li:2012bt}, and the quasipotential Bethe-Salpeter (BS) equation \cite{Song:2022svi,Song:2023vtu,Kong:2023dwz}.

In this work, we systematically investigate the existence of $S$-wave bound states composed of a heavy baryon and an antiheavy baryon or double heavy baryons in the BS equation approach within the ladder approximation and the instantaneous approximation for the kernel. Our model incorporates one $\textit{free}$ parameter, the cutoff parameter $\Lambda$, which is actually not entirely free as it governs the range of interaction and is directly related to the hadron size. Considering that the system may involve contributions from multiple exchange particles, with different interaction ranges, we reparametrize the cutoff parameter $\Lambda$ as $\Lambda = m + \alpha\Lambda_{\rm{QCD}}$ with $m$ being the mass of the exchange particle. This approach allows for different cutoffs for various exchange particles through the varying parameter $\alpha$ which is of order unity.

This work is organized as follows. After the introduction,
we present the formalism in Sec. \ref{Formalism}, which contains the
Lagrangians and the BS equations for the heavy baryonium and heavy dibaryon systems. In Sec. \ref{num results}, we show the numerical results for the the heavy baryonium and heavy dibaryon systems. Finally, Sec. \ref{sum and dis} provides a brief summary and discussion. The isospin conventions and the wave functions for the charmed baryonium and charmed dibaryon systems are given in Appendix.

\section{Formalism}
\label{Formalism}
To study whether the $S$-wave bound states of heavy baryonium and dibaryon exist, we fist construct the Lagrangians for heavy baryons and light mesons. Then the interaction kernels for the BS equations will be derived from the four-point Green’s function with the relevant Lagrangians.

\subsection{Effective chiral Lagrangians}
\label{ECL}
A heavy baryon contains a heavy quark and two light quarks, which will be refered to as a diquark in the following. Each light quark is in a triplet representation of the flavor SU(3), thus the diquark can form either an antisymmetric antitriplet or a symmetric sextet. The diquark in the flavor-antisymmetric antitriplet has spin 0, and the diquark in the flavor-symmetric sextet has spin 1. Considering a ground state heavy baryon, the diquark combined with the heavy quark can form an antitriplet baryon with spin-$\frac12$ $\left(B^{(Q)}_{\bar{3}}\right)$ and two sextet baryons with spin-$\frac12$ $\left(B^{(Q)}_6\right)$ and spin-$\frac32$ $\left(B^{(Q)\ast}_6\right)$, respectively. The heavy baryon matrices are
\begin{equation}
\label{Bc}
B^{(c)}_{\bar{3}}=\left(
\begin{array}{ccc}      0     &\Lambda_c^+&\Xi_c^+\\
                  -\Lambda_c^+&    0      &\Xi_c^0\\
                     -\Xi_c^+ &  -\Xi_c^0 &    0  \\
\end{array} \right),
B^{(c)}_{6}=\left(
\begin{array}{ccc}
\Sigma_c^{++}&\frac{1}{\sqrt{2}}\Sigma_c^+&\frac{1}{\sqrt{2}}\Xi_c^{'+}\\
\frac{1}{\sqrt{2}}\Sigma_c^+&    \Sigma_c^0      &\frac{1}{\sqrt{2}}\Xi_c^{'0}\\
\frac{1}{\sqrt{2}}\Xi_c^{'+}&\frac{1}{\sqrt{2}}\Xi_c^{'0}&\Omega_c^0  \\
\end{array} \right),
\end{equation}
\begin{equation}
\label{Bb}
B^{(b)}_{\bar{3}}=\left(
\begin{array}{ccc}      0     &\Lambda_b^0&\Xi_b^0\\
                  -\Lambda_b^0&    0      &\Xi_b^-\\
                     -\Xi_b^0 &  -\Xi_b^- &    0  \\
\end{array} \right),
B^{(b)}_{6}=\left(
\begin{array}{ccc}
\Sigma_b^{+}&\frac{1}{\sqrt{2}}\Sigma_b^0&\frac{1}{\sqrt{2}}\Xi_b^{'0}\\
\frac{1}{\sqrt{2}}\Sigma_b^0&    \Sigma_b^-      &\frac{1}{\sqrt{2}}\Xi_b^{'-}\\
\frac{1}{\sqrt{2}}\Xi_b^{'0}&\frac{1}{\sqrt{2}}\Xi_b^{'-}&\Omega_b^-  \\
\end{array} \right),
\end{equation}
and the matrices for $B^{(Q)\ast}_6$ are similar to those for $B^{(Q)}_6$.

For convenience, while performing chiral-loop calculations, the two sextet heavy baryons can be combined to a superfield,
\begin{equation}
\begin{split}
    S^\mu=&B_6^{\ast\mu}-\frac{1}{\sqrt{3}}(\gamma^\mu+v^\mu)\gamma_5B_6,\\
    \bar{S}^\mu=&\bar{B}_6^{\ast\mu}-\frac{1}{\sqrt{3}}(\gamma^\mu+v^\mu)\gamma_5\bar{B}_6,
\end{split}
\end{equation}
where $v^\mu$ is the velocity of the heavy baryon.

Then the general chiral Lagrangian for heavy baryons is \cite{Liu:2011xc,Cheng:1993kp}
\begin{equation}
    \mathcal{L}_B=\mathcal{L}_{\bar{3}}+\mathcal{L}_S+\mathcal{L}_{int},
\end{equation}
with
\begin{equation}
    \begin{split}\label{LB3}
        \mathcal{L}_{\bar{3}}=&\frac12\mathrm{tr}\left[\bar{B}_{\bar{3}}(i v\cdot D)B_{\bar{3}}\right]+i\beta_B\mathrm{tr}\left[\bar{B}_{\bar{3}}v^\mu(\mathcal{V}_\mu-\rho_\mu)B_{\bar{3}}\right]+\ell_B\mathrm{tr}\left[\bar{B}_{\bar{3}}\sigma B_{\bar{3}}\right],\\
    \end{split}
\end{equation}
\begin{equation}
    \begin{split}\label{LB6}
        \mathcal{L}_{S}=&-\mathrm{tr}\left[\bar{S}_\alpha(i v\cdot D-\Delta_B)S^\alpha\right]+\frac{3}{2}g_1(iv_\kappa)\epsilon^{\mu\nu\lambda\kappa}\mathrm{tr}\left[\bar{S}_\mu \mathcal{A}_\nu S_\lambda\right]\\
        &+i\beta_S\mathrm{tr}\left[\bar{S}_\mu v_\alpha(\mathcal{V}^\alpha-\rho^\alpha)S^\mu\right]+\lambda_S\mathrm{tr}\left[\bar{S}_\mu F^{\mu\nu}S_\nu\right]+\ell_S\mathrm{tr}\left[\bar{S}_\mu \sigma S^\mu\right],\\
    \end{split}
\end{equation}
\begin{equation}
    \begin{split}\label{LB}
       \mathcal{L}_{int} =&g_4\mathrm{tr}\left[\bar{S}_\mu \mathcal{A}^\mu B_{\bar{3}}\right]+i\lambda_I \epsilon^{\mu\nu\lambda\kappa}\mathrm{tr}\left[\bar{S}_\nu F^{\mu\nu}S_\nu\right]+\mathrm{H.c.},
    \end{split}
\end{equation}
where $D_\mu B=\partial_\mu B+\mathcal{V}_\mu B+B\mathcal{V}_\mu^T$, $\Delta_B=M_6-M_{\bar{3}}$ is the mass difference between the sextet and the antitriplet, $\mathcal{V}_\mu=\frac12\left(\xi^\dag\partial_\mu\xi+\xi\partial_\mu\xi^\dag\right)$ and $\mathcal{A}_\mu=\frac12\left(\xi^\dag\partial_\mu\xi-\xi\partial_\mu\xi^\dag\right)$ are the vector and axial vector fields, respectively, $F_{\mu\nu}=\partial_\mu\rho_\nu-\partial_\nu\rho_\mu+\left[\rho_\mu,\rho_\nu\right]$, $\xi=\mathrm{exp}[i{P}/f_\pi]$ and $\rho=ig_V/\sqrt{2}{V}$ with 
\begin{equation}
\label{pseudoscalar}
{P}=\left(
\begin{array}{ccc} \frac{\pi^0}{\sqrt{2}}+\frac{\eta}{\sqrt{6}} &\pi^+&K^+\\
                  \pi^-& -\frac{\pi^0}{\sqrt{2}}+\frac{\eta}{\sqrt{6}} &K^0\\
                     K^- &  \bar{K}^0 &    -\sqrt{\frac{2}{3}}\eta  \\
\end{array} \right),
\end{equation}
and
\begin{equation}
\label{vector}
{V}=\left(
\begin{array}{ccc} \frac{\omega}{\sqrt{2}}+\frac{\rho^0}{\sqrt{2}} &\rho^+&K^{\ast+}\\
                  \rho^-&\frac{\omega}{\sqrt{2}}-\frac{\rho^0}{\sqrt{2}}&K^{\ast0}\\
                     K^{\ast-} &  \bar{K}^{\ast+} &    \phi  \\
\end{array} \right),
\end{equation}
being the pseudoscalar and vector matrices, respectively. The phases of the fields $\bar{B}_{\bar{3}}$, $\bar{B}^{(\ast)}_6$, $\mathcal{V}_\mu^T$, and $\mathcal{A}_\mu^T$ can be fixed by the following charge conjugation convention:
\begin{equation}  \bar{B}_{\bar{3}}=\mathcal{C}B_{\bar{3}}\mathcal{C}^{-1},\quad\bar{B}^{(\ast)}_6=\mathcal{C}B^{(\ast)}_6\mathcal{C}^{-1},\quad \mathcal{V}_\mu^T=-\mathcal{C}\mathcal{V}_\mu\mathcal{C}^{-1},\quad \mathcal{A}_\mu^T=\mathcal{C}\mathcal{A}_\mu\mathcal{C}^{-1}.
\end{equation}

After expanding the effective Lagrangians in Eqs.(\ref{LB3})-(\ref{LB}) to the leading order of the light meson field, we can obtain the following effective interactions needed for our work:
\begin{equation}\label{Lagrangians}
    \begin{split}
        \mathcal{L}_{B_{\bar{3}}B_{\bar{3}}V}&=i\frac{\beta_Bg_V}{2\sqrt{2m_{\bar{B}_{\bar{3}}}m_{B_{\bar{3}}}}}\mathrm{tr}\left[\bar{B}_{\bar{3}}\overleftrightarrow{\partial}_\mu V^\mu B_{\bar{3}}\right],\\
        \mathcal{L}_{B_{\bar{3}}B_{\bar{3}}\sigma}&=\ell_B \mathrm{tr}\left[\bar{B}_{\bar{3}}\sigma B_{\bar{3}}\right],\\
        \mathcal{L}_{B_{6}B_{6}P}&=-\frac{g_1}{4f_\pi\sqrt{m_{\bar{B}_6}m_{B_6}}}\epsilon_{\mu\nu\lambda\kappa}\mathrm{tr}\left[\bar{B}_6\gamma^\mu\gamma^\lambda\overleftrightarrow{\partial}^\kappa \partial^\nu P B_6\right],\\
        \mathcal{L}_{B_{6}B_{6}V}&=-i\frac{\beta_Sg_V}{2\sqrt{2m_{\bar{B}_6}m_{B_6}}}\mathrm{tr}\left[\bar{B}_6\overleftrightarrow{\partial}_\nu V^\nu B_{6} \right]-i\frac{\lambda_Sg_V}{3\sqrt{2}}\mathrm{tr}\left[\bar{B}_6\gamma^\mu\left(\partial_\mu V_\nu-\partial_\nu V_\mu\right)\gamma^\nu B_{6}\right],\\
        \mathcal{L}_{B_{6}B_{6}\sigma}&=-\ell_S\mathrm{tr}\left[\bar{B}_6\sigma B_{6}\right],\\
    \end{split}
\end{equation}
where $v$ is replaced by $i\overleftrightarrow{\partial}/\left(2\sqrt{m_{\bar{B}}m_B}\right)$ and the pion decay constant is $f_\pi=132$ MeV. The values of relevant coupling constants are listed in Table \ref{Coupling constants} \cite{Song:2023vtu}.
\begin{table}[h]
\renewcommand{\arraystretch}{1.3}
\centering
\caption{Coupling constants.}
\begin{tabular*}{\textwidth}{@{\extracolsep{\fill}}ccccccc}
\hline
\hline 
 $g_V$   & $\beta_B$ & $\beta_S$     &  $\ell_B$   & $\ell_S$   & $g_1$   & $\lambda_S$ \\
  5.9    &   0.87    & $-2\beta_B$   &  $-3.1$     & $-2\ell_B$ & $-0.94$ & $3.31 \,\mathrm{GeV^{-1}}$ \\ 
\hline\hline
\end{tabular*}\label{Coupling constants}
\end{table}

\subsection{The BS equation for the heavy baryonium system}
 In this section, we will discuss the general BS formalism for the heavy baryonium composed of a heavy baryon and an anti-heavy baryon. In this case, the BS wave function is defined as: 
\begin{equation}
\begin{split}\label{BS wf baryonium}
\chi_P(x_1,x_2,P)_{\alpha\beta}&=\langle0|T\psi_\alpha(x_1)\bar{\psi}_\beta(x_2)|P\rangle,\\
&=e^{-iP X}\int\frac{d^4p}{(2\pi)^4}e^{-ip x}\chi_P(p)_{\alpha\beta},\\
\end{split}
\end{equation}
where $\alpha$ and $\beta$ are spinor indices, $\psi(x_1)$ and $\bar{\psi}(x_2)$ are
the field operators of heavy baryon and anti-heavy baryon, respectively, $P$ ($=Mv$) is the total momentum of the heavy baryonium and $v$ represents its velocity, $X=\lambda_1x_1-\lambda_2x_2$ and $x=x_1-x_2$ are the center-of-mass coordinate and the relative coordinate of the heavy baryonium, respectively, with $\lambda_{1(2)}=\frac{m_{1(2)}}{m_{1}+m_{2}}$, where $m_{1}$ and $m_{2}$ are the masses of heavy baryon and anti-heavy baryon, respectively, $p$ is the relative momentum of the heavy baryonium. The momenta of constituent particles can be expressed in terms of the relative momentum $p$ and the total momentum $P$ as $p_1=\lambda_1P+p$ and $p_2=\lambda_2P-p$, respectively.

The BS equation for the heavy baryonium can be written as
\begin{equation}\label{BS baryonium}
    \chi_P(p)=S(p_1)\int\frac{d^4q}{(2\pi)^4}\bar{K}(P,p,q)\chi_P(q)S(-p_2),
\end{equation}
where $\bar{K}(P,p,q)$ is the interaction kernel, can derived from the irreducible Feynman diagrams, $S(p_{1})$ and $S(-p_2)$ are the propagators of the heavy baryon and the anti-heavy baryon, respectively. For convenience, we define $p_l\equiv v\cdot p$ as the longitudinal projection of $p$ along $v$, and $p_t\equiv p-p_l v$ as the transverse component with respect to $v$.

In the leading of a $1/m_Q$ expansion, the propagators of the heavy baryon and the anti-heavy baryon can be expressed as:
\begin{equation}\label{propagator one}
       S(p_1) =i\frac{m_1(1+\slashed{v})}{2w_1(\lambda_1M+p_l-w_1+i\epsilon)},
\end{equation}
and
\begin{equation}\label{propagator two}
       S(p_2)=i\frac{m_2(1+\slashed{v})}{2w_2(\lambda_2M-p_l-w_2+i\epsilon)},
\end{equation}
where the energy $w_{1(2)}=\sqrt{m^2_{1(2)}-p_t^2}$, and $\epsilon$ is an infinitesimal parameter.

Substituting Eqs. (\ref{propagator one}) and (\ref{propagator two}) into Eq. (\ref{BS baryonium}), we obtain the following two constraint relations for the BS wave function $\chi_P(p)$:
\begin{equation}\label{con-rel1}
    \slashed{v}\chi_P(p)=\chi_P(p),
\end{equation}
\begin{equation}\label{con-rel2}
    \chi_P(p)\slashed{v}=-\chi_P(p).
\end{equation}

The $S$-wave heavy baryonium can have $J^{PC}=0^{-+}$ and $J^{PC}=1^{--}$ states. With the constraints imposed by parity and Lorentz transformations, the BS wave functions can be expressed as the following: 
\begin{equation}
    \chi_P(p)=\gamma_5f_1+\gamma_5\slashed{v}f_2+\gamma_5\slashed{p}_tf_3+(-i)\sigma_{\mu\nu}v^\mu p_t^\nu f_4,
\end{equation}
and 
\begin{equation}
   \begin{split}
    \chi_P^{(r)}(p)=&\Big[p_t^\rho g_1+\gamma_\mu\left(v^\mu p_t^\rho g_2+p_t^\mu p_t^\rho g_3+g^{\mu\rho}g_4\right)+\gamma_5\gamma_\mu\epsilon^{\mu\rho\alpha\beta}p_{t\alpha}v_\beta g_5\\
    &+\sigma_{\mu\nu}\left(p_t^\mu v^\nu p_t^\rho g_6+g^{\mu\rho}p_t^\nu g_7+g^{\mu\rho}v^\nu g_8\right)\Big]\epsilon_\rho^{(r)},
    \end{split}
\end{equation}
for the $J^{PC}=0^{-+}$ and $J^{PC}=1^{--}$ $S$-wave heavy baryonia, respectively, where $f_i$ $(i=1,...,4)$ and $g_j$ $(j=1,...,8)$ are the Lorentz-scalar functions of $p_t^2$ and $p_l$, and $\epsilon^{(r)}_\mu$ is the polarization vector of the vector heavy baryonium.

By applying the constraint relations (\ref{con-rel1}) and (\ref{con-rel2}), the BS wave functions for the $S$-wave pseudoscalar ($J^{PC}=0^{-+}$) and vector ($J^{PC}=1^{--}$) heavy baryonia can be simplified to the following forms, respectively:
\begin{equation}\label{sim BS PWF}
    \chi_P(p)=(1+\slashed{v})\gamma_5f_1,
\end{equation}
and
\begin{equation}\label{sim BS VWF}
    \chi^{(r)}_P(p)=(1+\slashed{v})\slashed{\epsilon}^{(r)}g_1.
\end{equation}

\subsection{The BS equation for the heavy dibaryon}
For the heavy dibaryon bound states composed of double heavy baryons, the general
form of the BS equationin in momentum space is: 
\begin{equation}\label{BS dibaryon}
    \chi_P(p)=S(p_1)\int\frac{d^4q}{(2\pi)^4}\bar{K}(P,p,q)\chi_P(q)S(p_2),
\end{equation}
with the BS wave function for the heavy dibaryon being defined as:
\begin{equation}\label{BS wf dibaryon}
\chi_P(x_1,x_2,P)_{\alpha\beta}=\langle0|T\psi_\alpha(x_1)\psi_\beta(x_2)|P\rangle.
\end{equation}
For convenience, we define a deformed BS wave function,
\begin{equation}\label{deformed BS wf dibaryon}
\tilde{\chi}_P(p)_{\alpha\beta}=\chi_P(p)_{\alpha\gamma}\mathcal{C}_{\gamma\beta}^{-1}=\left(\mathcal{C}\chi_P(p)\right)_{\alpha\beta},
\end{equation}
where $\mathcal{C}$ is the charge conjugation matrix. 

With this deformed BS wave function, the BS equation (\ref{BS dibaryon}) can be written in a more conventional matrix form
\begin{equation}\label{deformed BS dibaryon}
    \tilde{\chi}_P(p)^T=S(p_1)\int\frac{d^4q}{(2\pi)^4}\bar{K}(P,p,q)\tilde{\chi}_P(q)^TS(-p_2),
\end{equation}
where the superscript $``T"$ represents the transpose of the spinor index.

From the BS equations (\ref{BS baryonium}) and (\ref{deformed BS dibaryon}), we see that the BS wave function $\chi_P(p)$ in Eq. (\ref{BS wf baryonium}) for the heavy baryonium and the deformed BS wave function $\tilde{\chi}_P(p)$ in Eq. (\ref{deformed BS dibaryon}) for the heavy dibaryon satisfy the same equation. And the deformed BS wave function $\tilde{\chi}_P(p)$ have the same forms as given in Eqs. (\ref{sim BS PWF}) and (\ref{sim BS VWF}). 

To simplify the BS equations (\ref{BS baryonium}) and (\ref{deformed BS dibaryon}), we impose the so-called covariant instantaneous approximation in the kernel: $p_l=q_l$. In this approximation, the projection of the momentum of each constituent particle along the total momentum $P$ is not changed, i.e., the energy exchanged between the constituent particles of the binding system is neglected. This approximation is appropriate since we consider the binding energy of heavy baryonium and heavy dibaryon bound states to be very small compared to the masses of heavy baryons. Under this approximation, the kernel in the BS equation is reduced to $\bar{K}(P,p_t,q_t)$, which will be used in the following calculations.

After some algebra, we find that the BS scalar wave functions $f_1$ and $g_1$ satisfy the same integral equation as follows (in the following we will use $f$ uniformly):
\begin{equation}\label{BS scalar WF}
    f(p)=\frac{-m_1m_2}{2w_1w_2(\lambda_1M+p_l-w_1+i\epsilon)(\lambda_2M-p_l-w_2+i\epsilon)}\int\frac{d^4q}{(2\pi)^4}\bar{K}(P,p_t,q_t)f(q).
\end{equation}
We integrate both sides of the above equation with respect to $p_l$ to obtain:
\begin{equation}\label{3D-BS scalar WF}
    \tilde{f}(p_t)=\frac{-im_1m_2}{w_1w_2(M-w_1-w_2)}\int\frac{d^3q_t}{(2\pi)^3}\bar{K}(P,p_t,q_t)\tilde{f}(q_t),
\end{equation}
where we have defined $\tilde{f}(p_t)=\int dp_l f(p)$.

Based on the effective Lagrangians in Eq.(\ref{Lagrangians}), the lowest-order interaction kernel can be derived as follows:
\begin{equation}
    \begin{split}
        \bar{K}_{B_{\bar{3}}\bar{B}_{\bar{3}}}^V(P,p,q)=&c_I\left(\frac{g_V\beta_B}{2\sqrt{2m_{B_{\bar{3}}}m_{\bar{B}_{\bar{3}}}}}\right)^2(p_1+q_1)_\mu(p_2+q_2)_\nu\Delta^{\mu\nu}_V(k),\\
        \bar{K}_{B_{\bar{3}}\bar{B}_{\bar{3}}}^\sigma(P,p,q)=&-c_I\ell_B^2\Delta_\sigma(k),\\
        \bar{K}_{B_{6}\bar{B}_{6}}^P(P,p,q)=&-c_I\left(\frac{g_1}{4f_\pi\sqrt{m_{B_{6}}m_{\bar{B}_{6}}}}\right)^2\epsilon_{\mu\nu\lambda\kappa}\epsilon_{\alpha\beta\tau\delta}\gamma^\mu\gamma^\lambda\gamma^\alpha\gamma^\tau(p_1+q_1)^\kappa(p_2+q_2)^\delta k^\nu k^\beta \Delta_P(k),\\
        \bar{K}_{B_{6}\bar{B}_{6}}^V(P,p,q)=&c_I\left[\frac{\beta_Sg_V}{2\sqrt{2m_{B_{6}}m_{\bar{B}_{6}}}}\left(p_1+q_1\right)^\mu g_{\mu\tau}-\frac{\lambda_Sg_V}{3\sqrt{2}}\gamma^\mu\gamma^\nu\left(k_\mu g_{\nu\tau}-k_\nu g_{\mu\tau}\right)\right]\\
        &\times\left[\frac{\beta_Sg_V}{2\sqrt{2m_{B_{6}}m_{\bar{B}_{6}}}}\left(p_2+q_2\right)^\alpha g_{\alpha\kappa}-\frac{\lambda_Sg_V}{3\sqrt{2}}\gamma^\alpha\gamma^\beta\left(k_\alpha g_{\beta\kappa}-k_\beta g_{\alpha\kappa}\right)\right]\Delta_V^{\tau\kappa}(k),\\
        \bar{K}_{B_{6}\bar{B}_{6}}^\sigma(P,p,q)=&-c_I\ell_S^2\Delta_\sigma(k),
    \end{split}
\end{equation}
where $\Delta^{\mu\nu}_V(k)$, $\Delta_P(k)$, and $\Delta^\sigma(k)$ are the propagators of the exchanged vector, pseudoscalar and $\sigma$ mesons, respectively, $k$ represents the momentum of the exchanged meson, and $c_I$ is the isospin coefficient, given in Table \ref{Isospin factors}. In our model, the BS wave function depends only on the isospin $I$ but not on its
component $I_3$ because we consider only strong interactions that preserve the isospin symmetry.

\begin{table}[h]
\renewcommand{\arraystretch}{1.3}
\centering
\caption{Isospin factors. The values outside and inside brackets are for heavy baryonium and heavy dibaryon systems, respectively.}
\begin{tabular*}{\textwidth}{@{\extracolsep{\fill}}lccccccccc}
\hline
\hline 
 & $\Lambda_c\bar{\Lambda}_c(\Lambda_c)$ &\multicolumn{2}{c}{$\Xi_c\bar{\Xi}_c(\Xi_c)$} & \multicolumn{3}{c}{$\Sigma_c\bar{\Sigma}_c(\Sigma_c)$}  &\multicolumn{2}{c}{$\Xi'_c\bar{\Xi}'_c(\Xi'_c)$}& $\Omega_c\bar{\Omega}_c(\Omega_c)$\\
 \hline
 $I$                 & 0     & 0          &1          &0          &1           &2               &0           &1           &0    \\
$\mathcal{C}_\pi$    &       &            &           &1[-1]      &$\frac12[-\frac12]$ &$-\frac12[\frac12]$&$\frac38[-\frac38]$ &$-\frac18[\frac18]$ &     \\
$\mathcal{C}_\eta$   &       &            &           &$\frac16[\frac16]$ &$\frac16[\frac16]$  &$\frac16[\frac16]$ &$\frac{1}{24}[\frac{1}{24}]$&$\frac{1}{24}[\frac{1}{24}]$&$\frac23[\frac23]$\\ 
$\mathcal{C}_\rho$   &       &$-\frac32[-\frac32]$&$\frac12[\frac12]$ &$-1[-1]$   &$-\frac12[-\frac12]$&$\frac12[\frac12]$ &$-\frac38[-\frac38]$&$\frac18[\frac18]$  &     \\ 
$\mathcal{C}_\omega$ &$-2[2]$&$-\frac12[\frac12]$ &$-\frac12[\frac12]$&$-\frac12[\frac12]$&$-\frac12[\frac12]$ &$-\frac12[\frac12]$&$-\frac18[\frac18]$ &$-\frac18[\frac18]$ &     \\ 
$\mathcal{C}_\phi$   &       &$-1[1]$     &$-1[1]$    &           &            &           &$-\frac14[\frac14]$ &$-\frac14[\frac14]$ & $-1[1]$ \\ 
$\mathcal{C}_\sigma$ & 4[4]  & 4[4]       &4[4]       &1[1]       & 1[1]       &1[1]       &1[1]        &1[1]        & 1[1]   \\ 
\hline\hline
\end{tabular*}\label{Isospin factors}
\end{table}

To account for the structure and finite size effects of the interacting hadrons, it is necessary to introduce the form factor at the vertices. For $t$-channel vertices, we use the monopole form factor:
\begin{equation}
 \begin{split}
    F_M(k^2)=&\frac{\Lambda^2-m^2}{\Lambda^2-k^2},\\
 \end{split}   
\end{equation}
where $m$ and $\Lambda$ represent the mass and cutoff parameter of the exchanged meson, respectively. Since the heavy baryonium and heavy dibaryon systems can have interaction by exchanging multiple particles, different masses of exchanged particles correspond to different interaction ranges and, consequently, different cutoff parameters. Thus, we further reparameterize the cutoff $\Lambda$ as $\Lambda=m+\alpha \Lambda_{\rm{QCD}}$ with $\Lambda_{\rm{QCD}}$ = 220 MeV, where the parameter $\alpha$ is of order one. The value of $\alpha$ depends on the exchanged and external particles involved in the strong interaction vertex and cannot be obtained from the first principle.

\section{Numerical results}
\label{num results}
In the numerical calculations, we first present the masses of the relevant mesons and heavy baryons in Table \ref{masses} \cite{ParticleDataGroup:2024prd}, which are essential for investigating whether heavy baryonium and heavy dibaryon systems can exist as bound states. In our model, we have two parameters, the cutoff $\Lambda$ and the bounding energy $E_b$. The cutoff $\Lambda$ is reparameterized as a variable, $\alpha$, with $\Lambda=m+\alpha\Lambda_{\rm{QCD}}$. The parameter $\alpha$ is not a completely free parameter, since its varying range is related to the sizes of hadrons \cite{Chen:2017vai}. Based on the experience with deuteron, the parameter $\alpha$ is typically of order unity. The other parameter $E_b$ (defined as $E_b= m_1+m_2-M$, where we consider the heavy baryonium and dibaryon systems as shallow bound states with $E_b$ ranging from 0 to 50 MeV), is dependent on the value of the parameter $\alpha$, and is therefore not absolutely determined. In this work, we allow the parameter $\alpha$ to vary over a wide range (0.3–8) to search for possible solutions in the heavy baryonium and dibaryon systems.

To solve the three-dimensional integral BS equation (\ref{3D-BS scalar WF}), we fist simplify it to a one-dimensional integral equation by completing the azimuthal integration. This one-dimensional integral BS equation is further discretised into a matrix eigenvalue equation by the Gaussian quadrature method. By solving the eigenvalue equation, we can find the possible bound states of the heavy baryonium and heavy dibaryon systems depending on the parameter $\alpha$.

\begin{table}[h]
\renewcommand{\arraystretch}{1.3}
\centering
\caption{Masses (in MeV) of mesons and heavy baryons. The bottom baryons $\Sigma_b^0$ and $\Xi_b^{'0}$ have not been observed experimentally, thus, we use $m_{\Sigma_b^0}=\frac{1}{2}\left(m_{\Sigma_b^+}+m_{\Sigma_b^-}\right)$ and $m_{\Xi_b^{'0}}=m_{\Xi_b^{'-}}$.}
\begin{tabular*}{\textwidth}{@{\extracolsep{\fill}}ccccccccc}
\hline
\hline 
 $m_{\pi^\pm}$ & $m_{\pi^0}$ & $m_\eta$ & $m_\sigma$  &  $m_\rho$  & $m_\omega$ & $m_\phi$    \\
 \hline
  139.57 & 134.977  & 547.862  & 500         &  775.26    & 782.66     & 1019.461  \\ 
  \hline\hline
  $m_{\Lambda_c}$  &  $m_{\Xi_c^+}$ & $m_{\Xi_c^0}$ & $m_{\Sigma_c^{++}}$ & $m_{\Sigma_c^{+}}$ & $m_{\Sigma_c^{0}}$ & $m_{\Xi_c^{'+}}$ & $m_{\Xi_c^{'0}}$ & $m_{\Omega_c^{0}}$\\
  \hline
  2286.46 & 2467.71 & 2470.44 & 2453.97 & 2452.65 & 2453.75 & 2467.71 & 2470.44 & 2695.2 \\
\hline\hline
$m_{\Lambda_b}$  &  $m_{\Xi_b^0}$ & $m_{\Xi_b^-}$ & $m_{\Sigma_b^{+}}$ & $m_{\Sigma_b^{0}}$ & $m_{\Sigma_b^{-}}$ & $ m_{\Xi_b^{'0}}$ & $m_{\Xi_b^{'-}}$ & $m_{\Omega_b^{-}}$\\
  \hline
  5619.60 & 5791.9 & 5797.0 & 5810.56 & 5813.1 & 5815.64 & 5935.1 & 5935.1 & 6045.8 \\
\hline\hline
\end{tabular*}\label{masses}
\end{table}

\subsection{The results of charmed baryonium and charmed dibaryon systems}
\begin{figure}[htb]
\centering
\subfigure[]{
\includegraphics[width=5cm]{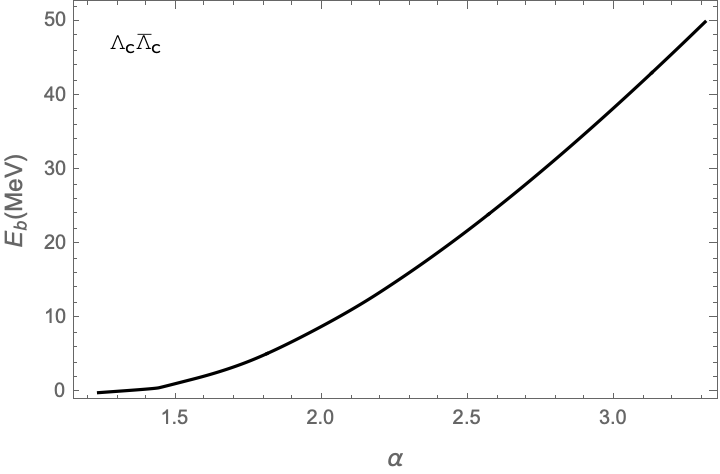}
}
\,
\subfigure[]{
\includegraphics[width=5cm]{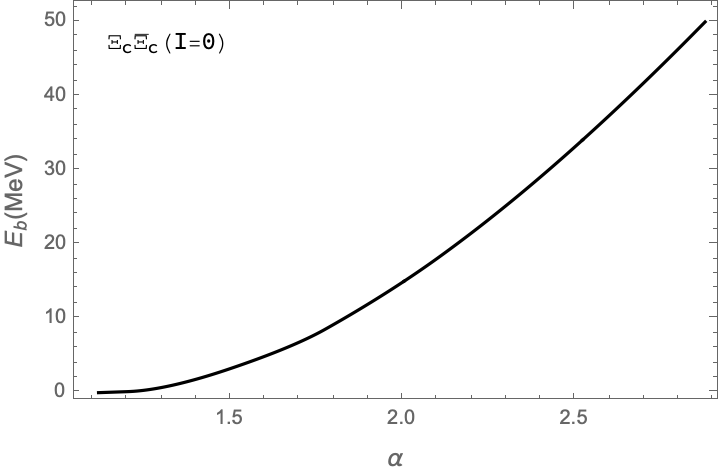}
}
\,
\subfigure[]{
\includegraphics[width=5cm]{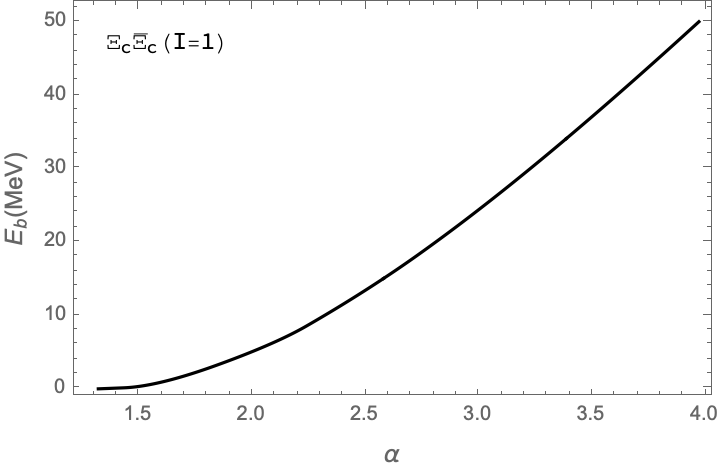}
}
\,
\subfigure[]{
\includegraphics[width=5cm]{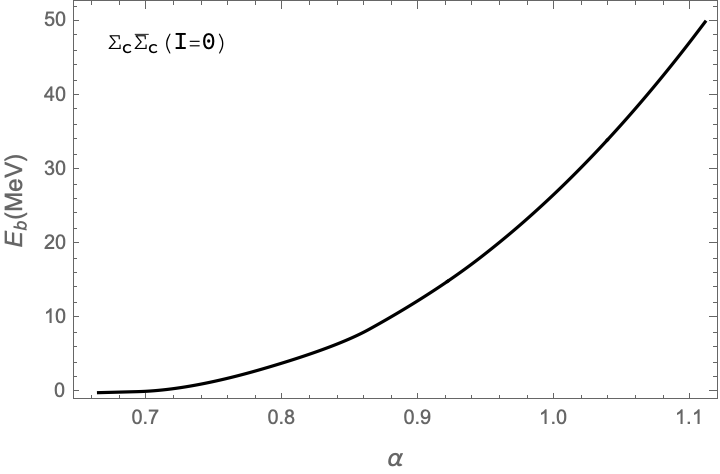}
}
\,
\subfigure[]{
\includegraphics[width=5cm]{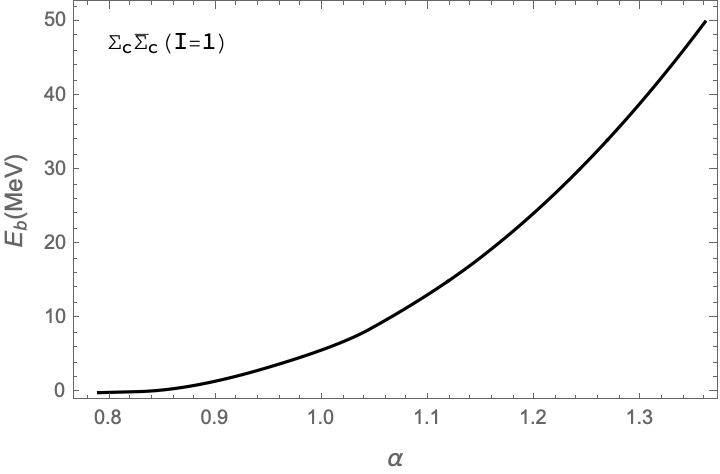}
}
\,
\subfigure[]{
\includegraphics[width=5cm]{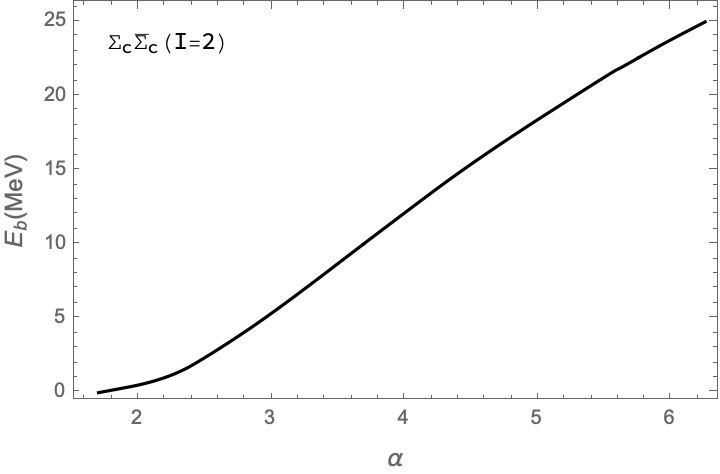}
}
\,
\subfigure[]{
\includegraphics[width=5cm]{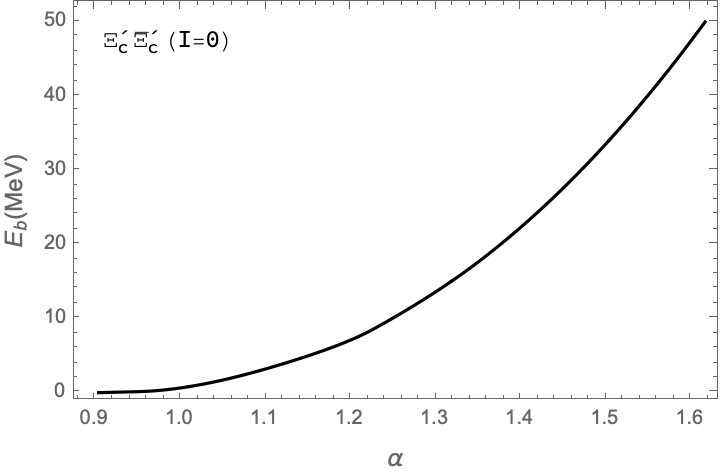}
}
\,
\subfigure[]{
\includegraphics[width=5cm]{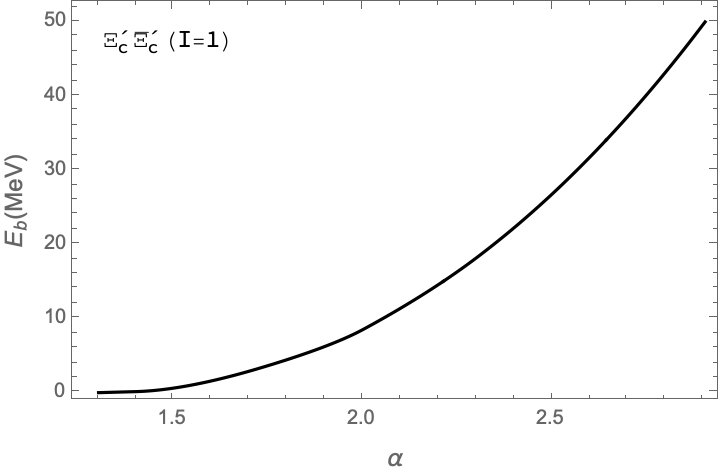}
}
\,
\subfigure[]{
\includegraphics[width=5cm]{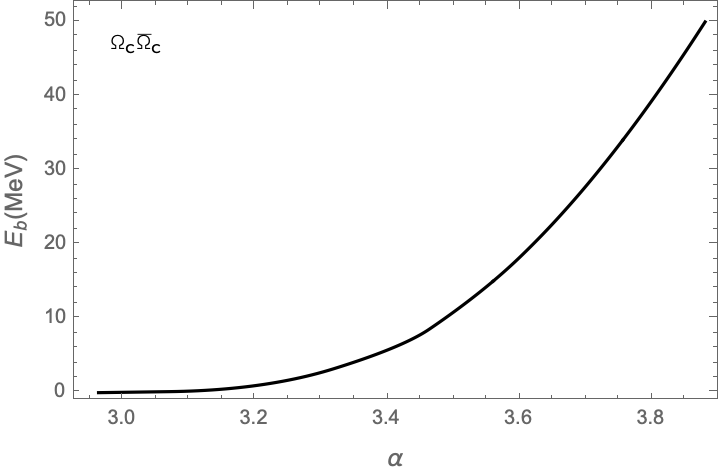}
}
\caption{Values of $\alpha$ and $E_b$ for the possible bound states of charmed baryonium systems.}
\label{Charmed baryonium}
\end{figure}

\begin{figure}[htb]
\centering
\subfigure[]{
\includegraphics[width=5cm]{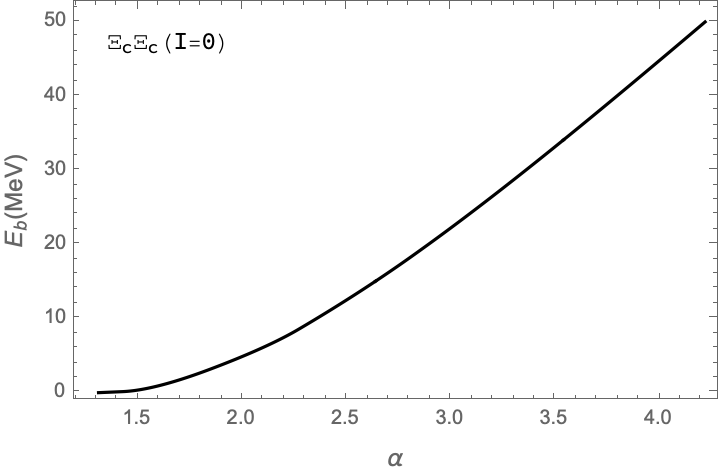}
}
\,
\subfigure[]{
\includegraphics[width=5cm]{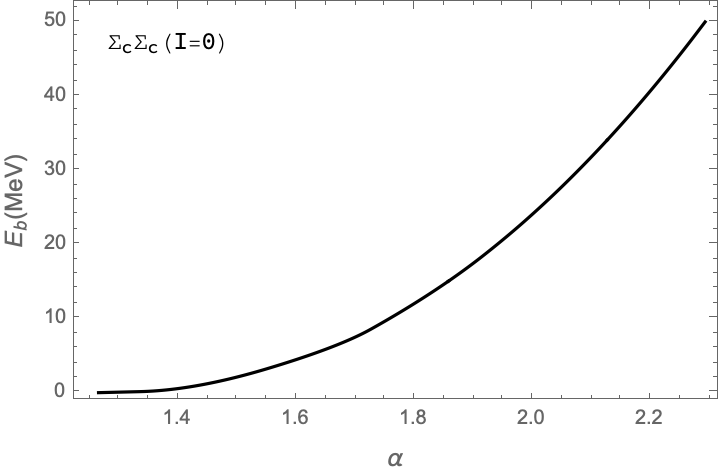}
}
\,
\subfigure[]{
\includegraphics[width=5cm]{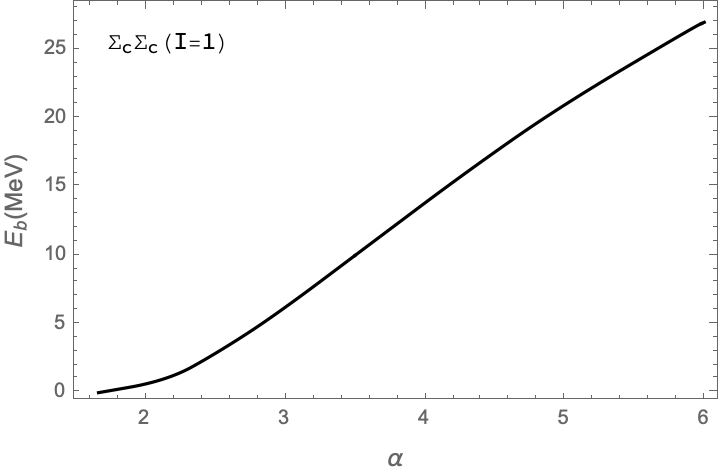}
}
\caption{Values of $\alpha$ and $E_b$ for the possible bound states of charmed dibaryon systems.}
\label{Charmed dibaryon}
\end{figure}

The results for the possible bound states of the charmed baryonium and charmed dibaryon systems are shown in Figs. \ref{Charmed baryonium} and \ref{Charmed dibaryon}, respectively. Our research indicates that all the charmed baryonium systems, specifically $\Lambda_c\bar{\Lambda}_c$, $\Xi_c\bar{\Xi}_c$, $\Sigma_c\bar{\Sigma}_c$, $\Xi'_c\bar{\Xi}'_c$, and $\Omega_c\bar{\Omega}_c$, can exist as bound states. Among the charmed dibaryon systems, only the $\Xi_c\Xi_c$ system with isospin $I=0$ and the $\Sigma_c\Sigma_c$ system with isospin $I=0$ and $I=1$ can exist as bound states. 

For the $\Lambda_c\bar{\Lambda}_c (\Lambda_c)$ system, since $\Lambda_c$ is an isoscalar state, the interaction kernel arises from the exchanges of $\omega$ and $\sigma$ mesons. Both $\omega$ and $\sigma$ mesons induce attractive interaction in the $\Lambda_c\bar{\Lambda}_c$ system, allowing it to form a bound state in our model. The values of the parameter $\alpha$ along with the corresponding binding energy $E_b$ are displayed in Fig. \ref{Charmed baryonium}(a). It is also found that this system can exist as a bound state in various models \cite{Lee:2011rka,Chen:2017vai,Lu:2017dvm,Chen:2011cta,Chen:2013sba,Dong:2021juy,Yu:2021lmb,Song:2022svi}. However, the results of Refs. \cite{Lu:2017dvm,Chen:2011cta,Chen:2013sba,Chen:2017vai,Lee:2011rka} show that the binding energy is sensitive to the cutoff $\Lambda$. 

For the $\Lambda_c\Lambda_c$ system, the interaction contributed from $\omega$ is repulsive and that from $\sigma$ is attractive. Our result indicates that the $\Lambda_c\Lambda_c$ system cannot form a bound state which is consistent with the results in Refs. \cite{Huang:2013rla,Carames:2015sya,Lee:2011rka,Dong:2021bvy} that the $\Lambda_c\Lambda_c$ system can not be a bound state by itself. However, the coupling of the $\Lambda_c\Lambda_c$ to the strongly attractive $\Sigma^{(\ast)}_c\Sigma^{(\ast)}_c$ system may lead to a state below the $\Lambda_c\Lambda_c$ threshold \cite{Meguro:2011nr,Li:2012bt,Huang:2013rla}. On the contrary, in Refs. \cite{Song:2022svi,Yu:2021lmb,Lu:2017dvm,Chen:2017vai}, it is pointed out that the single channel $\Lambda_c\Lambda_c$ can form a bound state. This discrepancy arises from the fact that in these models, the attractive contribution from the $\sigma$ meson is stronger than the repulsive contribution from the $\omega$ meosn. However, in our model, we find that even only considering the contribution from the $\sigma$ meson is not sufficient for the $\Lambda_c\Lambda_c$ system to form a bound state.

The $\Xi_c$ baryon contains a strange quark and has an isospin of 1/2. Therefore, the $\Xi_c\bar{\Xi}_c (\Xi_c)$ system can have isospins of both $I=0$ and $I=1$ and the interaction kernel can arise from the exchanges of $\rho$, $\omega$, $\phi$, and $\sigma$. For the $\Xi_c\bar{\Xi}_c$ system with $I=0$, the exchanges of $\rho$, $\omega$, $\phi$, and $\sigma$ mesons all induce attractive interaction. The $\Xi_c\bar{\Xi}_c$ system with $I=0$ can form a bound state with a binding energy in the range from 0 to 50 MeV when the parameter $\alpha$ ranges from 1.12 to 2.88, which is presented in Fig. \ref{Charmed baryonium}(b). In the $\Xi_c\bar{\Xi}_c$ system with $I=1$, the interaction magnitudes due to $\rho$ and $\omega$ are the same, but $\rho$ provides a repulsive contribution, thus their contributions almost cancel each other considering the similar masses of $\rho$ and $\omega$. Then the $\Xi_c\bar{\Xi}_c$ system with $I=1$ is able to exist as a bound state with a binding energy in the range of 0 to 50 MeV when the parameter $\alpha$ ranges from  1.32 to 3.97. The relevant results are presented in Fig. \ref{Charmed baryonium}(c). That the system $\Xi_c\bar{\Xi}_c$ with $I=0$ and $I=1$ can exist as a bound state is also supported by Refs. \cite{Lee:2011rka,Dong:2021juy,Song:2023vtu}. It is worth mentioning that in the lattice QCD \cite{Liu:2022gxf} and the chromomagnetic interaction model \cite{Liu:2021gva} it is found that the masses of the hidden-charm and hidden-strange hexaquarks are below the $\Xi_c\bar{\Xi}_c$ threshold by 700-1000 MeV, which cannot be obtained within a reasonable range of the parameter $\alpha$ in our model.

For the $\Xi_c\Xi_c$ system, only the isospin $I=0$ configuration can exist as a bound state, as depicted in Fig. \ref{Charmed dibaryon}(a). In the $\Xi_c\Xi_c$ system with $I=1$, contributions from the vector mesons $\rho$, $\omega$, and $\phi$ are repulsive, while that from the  $\sigma$ meson is attractive but insufficient to form a bound state in our model. However, the $\Xi_c\Xi_c$ system with $I=1$ could be a loosely bound state with a binding energy of only a few hundred keV within the one-boson-exchange model \cite{Lee:2011rka}, and it could be a deeply bound state within the quasipotential BS equation framework \cite{Song:2023vtu}. In our model, unlike the $\Lambda_c\Lambda_c$ system, the $\Xi_c\Xi_c$ system can form a bound state when only considering the $\sigma$ meson exchange due to the greater mass of the $\Xi_c$ compared with the $\Lambda_c$. However, given that the total contributions of the vector mesons in the $I=1$ $\Xi_c\Xi_c$ system are repulsive, it is quite inconceivable that this system could form a bound state.

For the $\Sigma_c\bar{\Sigma}_c(\Sigma_c)$ system, the interaction kernels are induced by the exchange of the pseudoscalar mesons $\pi$ and $\eta$, the vector mesons $\rho$ and $\omega$, and the scalar meson $\sigma$. In the $\Sigma_c\bar{\Sigma}_c$ system, our results indicate that the isospin states with $I = 0$, $1$, and $2$ can all exist as bound states, consistent with Refs. \cite{Lee:2011rka, Song:2022svi}. However, in our model, the $\Sigma_c\bar{\Sigma}_c$ system with $I = 2$ is a loosely bound state with the binding energy very sensitive to the parameter $\alpha$ compared with the isospin states $I = 0$ and $1$, as shown in Figs. \ref{Charmed baryonium}(d)-\ref{Charmed baryonium}(f). This sensitivity is due to the contributions from $\rho$ and $\omega$ mesons nearly canceling out, while the attraction from the $\pi$ meson accounts for the long-range interaction. In contrast, Ref. \cite{Lee:2011rka} reports a binding energy of 149.66 MeV for this state with a cutoff $\Lambda = 0.95$.

In the $\Sigma_c\Sigma_c$ system, the isospin states with $I=0$ and $I=1$ can form bound states, as presented in Figs. \ref{Charmed dibaryon}(b) and \ref{Charmed dibaryon}(c). Specifically, the $\Sigma_c\Sigma_c$ system with $I=1$ is a loosely bound state, consistent with the one-boson-exchange model \cite{Ling:2021asz} and the chiral effective theory \cite{Chen:2022iil}. The binding energy $E_b$ of the $\Sigma_c\Sigma_c$ system with $I=1$ is very sensitive to the parameter $\alpha$. In our model, contributions from pseudoscalar mesons $\pi$ and $\eta$, and vector mesons $\rho$ and $\omega$ are repulsive. Only the $\sigma$ meson provides an attractive force, which is insufficient to form a bound state for the $I=2$ $\Sigma_c\Sigma_c$ system. In contrast, Ref. \cite{Song:2022svi} suggests that the $\Sigma_c\Sigma_c$ system with $I=2$ can form a bound state.

For the $\Xi'_c\bar{\Xi}'_c(\Xi'_c)$ system, only the charmed baryonium $\Xi'_c\bar{\Xi}'_c$ system can exist as bound state, with results presented in Figs. \ref{Charmed baryonium}(g) and \ref{Charmed baryonium}(h). The charmed dibaryon $\Xi'_c\Xi'_c$ system cannot exist as a bound state in our model, which is inconsistent with Refs. \cite{Lee:2011rka} and \cite{Song:2023vtu}. According to Ref. \cite{Lee:2011rka}, as the root mean square (rms) radius increases, the vector mesons that originally provided repulsive contributions become attractive, allowing the $\Xi'_c\Xi'_c$ system to exist as a loosely bound state. For the $\Xi'_c\Xi'_c$ system with $I=0$, contributions from $\eta$, $\rho$, and $\sigma$ are attractive, while contributions from $\pi$, $\omega$, and $\phi$ are repulsive, but no bound state is found. Thus, except for the scalar $\sigma$ meson, which provides an attractive contribution, all other particles contribute repulsive forces to the $\Xi'_c\Xi'_c$ system with $I=1$, so that it cannot exist as a bound state in our model.

For the $\Omega_c\bar{\Omega}_c(\Omega_c)$ system, only the $\Omega_c\bar{\Omega}_c$ can exist as a bound state. In the $\Omega_c\Omega_c$ system, the $\rho$ and $\eta$ provide repulsive contributions, and the attractive contribution from $\sigma$ exchange alone is insufficient to form a bound state. However, Ref. \cite{Lee:2011rka} suggests that the $\Omega_c\Omega_c$ system can exist as a loosely bound state, because the repulsive contributions of $\eta$ and $\phi$ decrease rapidly as the rms radius increases, making the attraction contribution provided by $\sigma$ greater than the repulsion.

\subsection{The results of bottom baryonium and bottom dibaryon systems}

\begin{figure}[htb]
\centering
\subfigure[]{
\includegraphics[width=5cm]{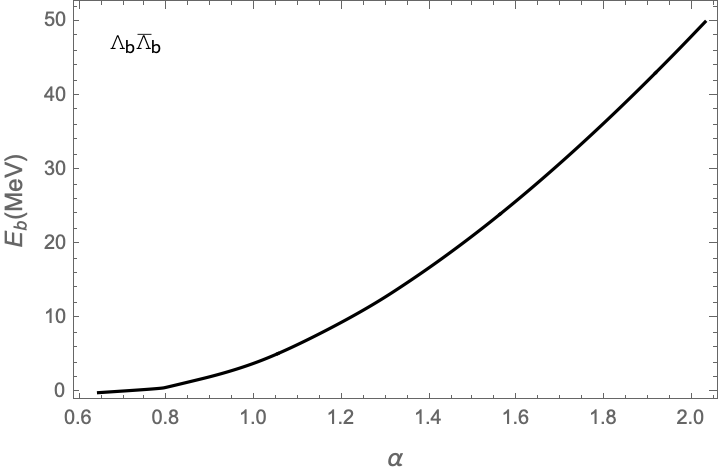}
}
\,
\subfigure[]{
\includegraphics[width=5cm]{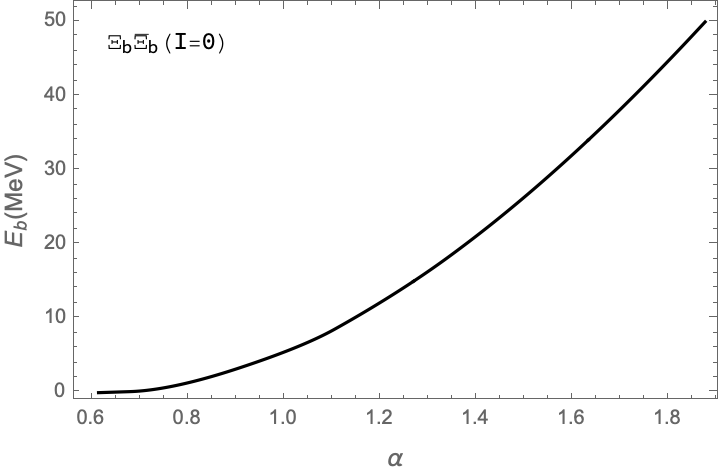}
}
\,
\subfigure[]{
\includegraphics[width=5cm]{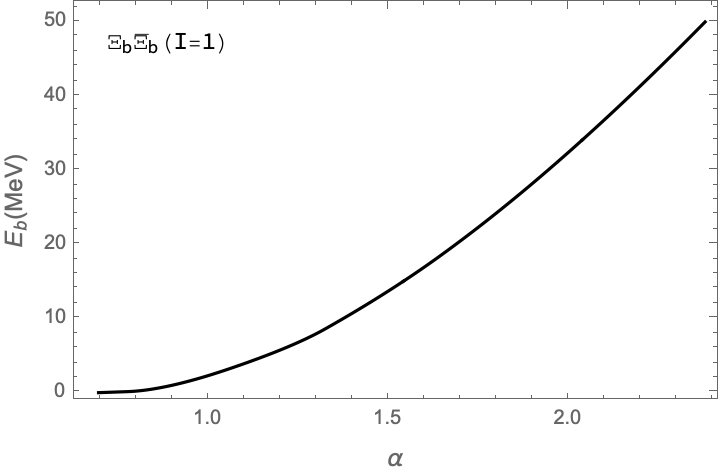}
}
\,
\subfigure[]{
\includegraphics[width=5cm]{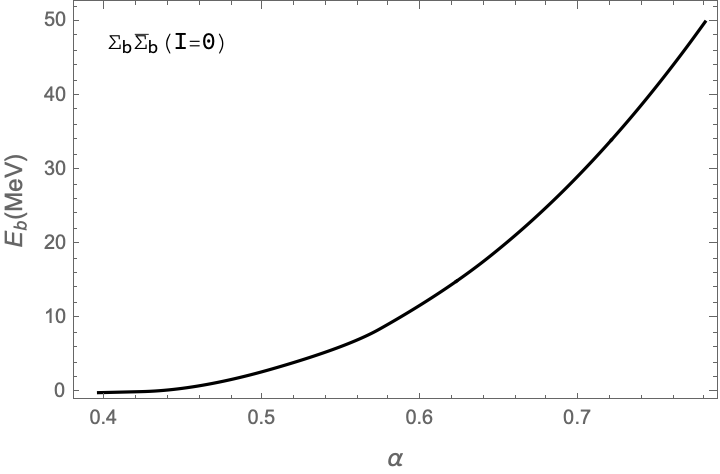}
}
\,
\subfigure[]{
\includegraphics[width=5cm]{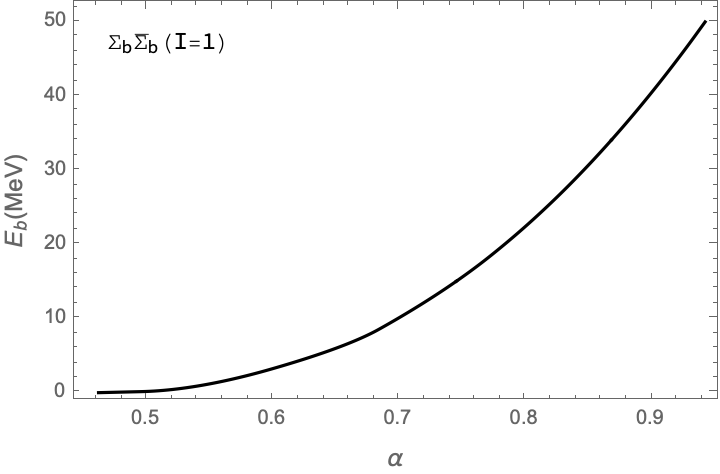}
}
\,
\subfigure[]{
\includegraphics[width=5cm]{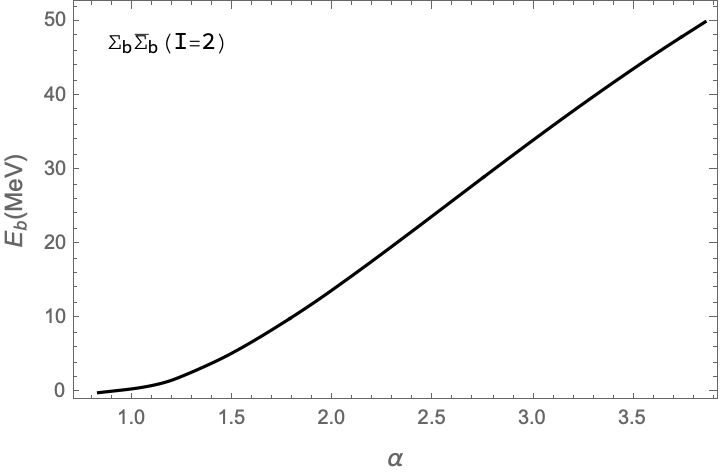}
}
\,
\subfigure[]{
\includegraphics[width=5cm]{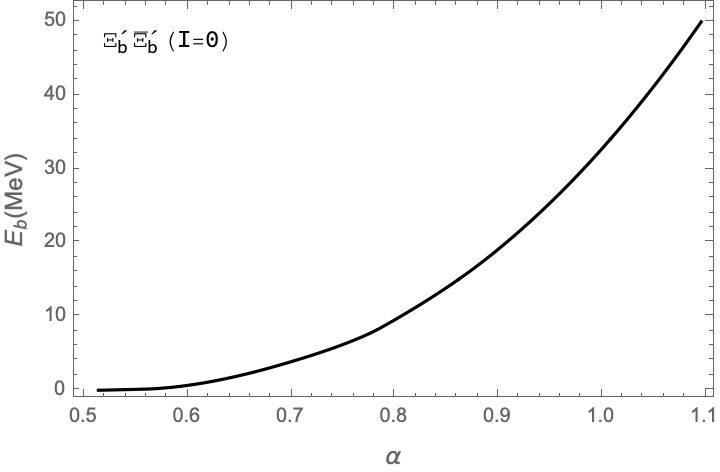}
}
\,
\subfigure[]{
\includegraphics[width=5cm]{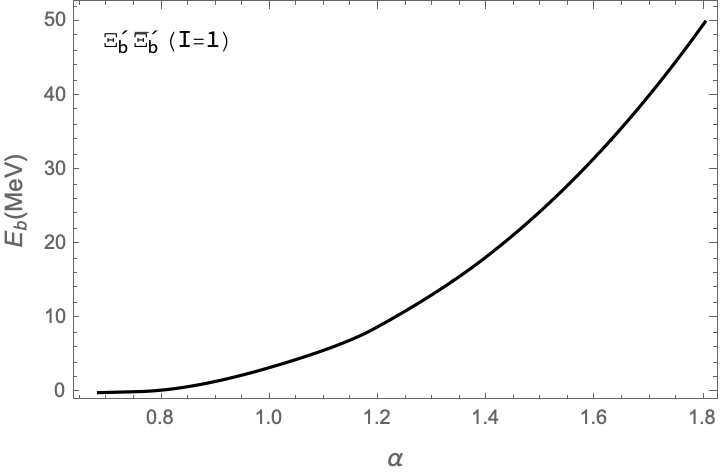}
}
\,
\subfigure[]{
\includegraphics[width=5cm]{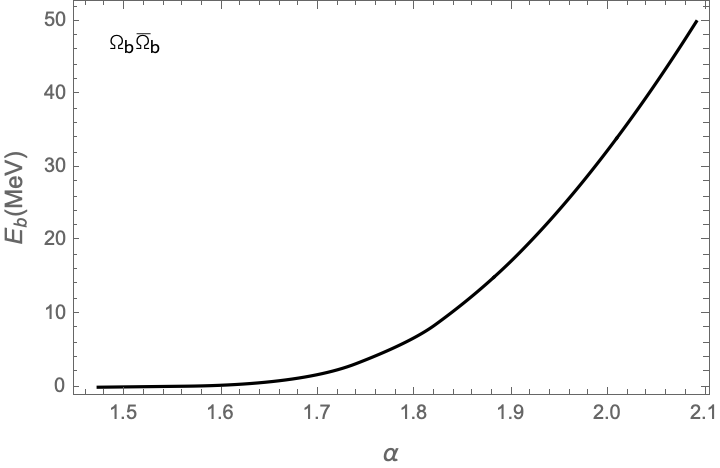}
}
\caption{Values of $\alpha$ and $E_b$ for the possible bound states of bottom baryonium systems.}
\label{Bottom baryonium}
\end{figure}

\begin{figure}[htb]
\centering
\subfigure[]{
\includegraphics[width=5cm]{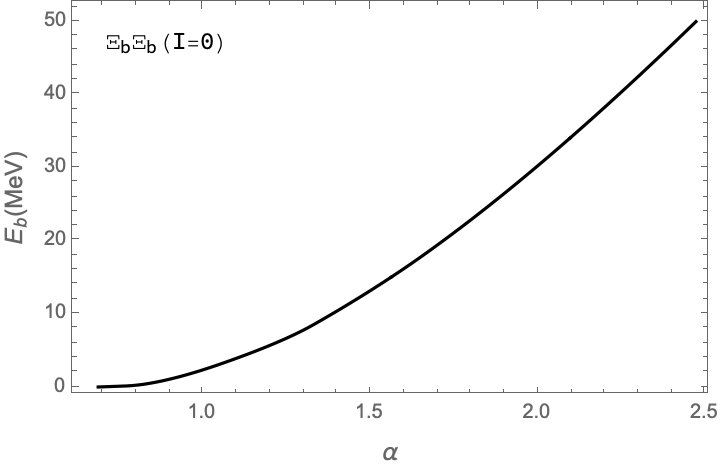}
}
\,
\subfigure[]{
\includegraphics[width=5cm]{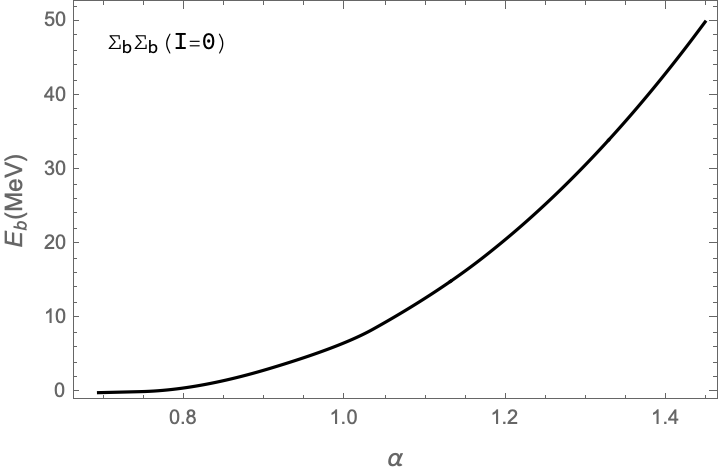}
}
\,
\subfigure[]{
\includegraphics[width=5cm]{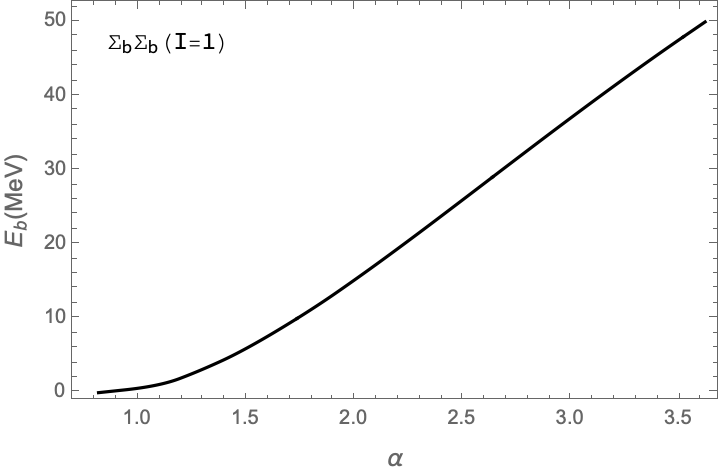}
}
\caption{Values of $\alpha$ and $E_b$ for the possible bound states of bottom dibaryon systems.}
\label{Bottom dibaryon}
\end{figure}

The interaction kernels in the bottom sector are the same as those in the charm sector. Thus, similar to the charmed baryonium and charmed dibaryon systems, all the bottom baryonium systems, such as $\Lambda_b\bar{\Lambda}_b$, $\Xi_b\bar{\Xi}_b$, $\Sigma_b\bar{\Sigma}_b$, and $\Omega_b\bar{\Omega}_b$, as well as the bottom dibaryon systems $\Xi_b\Xi_b$ with isospin $I=0$ and $\Sigma_b\Sigma_b$ with isospin $I=0$ and $I=1$, can exist as bound states. The results for the parameter $\alpha$ and the corresponding binding energy $E_b$ are displayed in Figs. \ref{Bottom baryonium} and \ref{Bottom dibaryon}. Because of the much heavier reduced masses of hidden-bottom systems, it is easier to form bound states than in charmed systems. Therefore, for the same binding energy, the bottom region corresponds to a smaller parameter $\alpha$.

Similar to the $\Sigma_c\bar{\Sigma}_c$ system with $I=2$ and the $\Sigma_c\Sigma_c$ system with $I=1$ in the charm region, the $I=2$ $\Sigma_b\bar{\Sigma}_b$ and $I=1$ $\Sigma_b\Sigma_b$ systems show the binding energies being very sensitive to the parameter $\alpha$. These two systems are also unable to bind very deeply, with the corresponding maximum binding energies of 70 MeV ($\alpha = 5.32$) and 62 MeV ($\alpha = 4.28$), respectively. To reasonably apply the instantaneous approximation in solving the BS equation (\ref{BS scalar WF}), we choose a maximum binding energy of 50 MeV. Therefore, the results for binding energies larger than 50 MeV are not shown in Figs. \ref{Charmed baryonium}, \ref{Charmed dibaryon}, \ref{Bottom baryonium}, and \ref{Bottom dibaryon}.

\section{Summary and Discussion}
\label{sum and dis}

In this work, we utilized the BS equation to systematically study whether heavy baryonium and heavy dibaryon systems can exist as bound states. Our research indicates that all the heavy baryonium systems, including $\Lambda_Q\bar{\Lambda}_Q$, $\Xi_Q\bar{\Xi}_Q$, $\Sigma_Q\bar{\Sigma}_Q$, $\Xi'_Q\bar{\Xi}'_Q$, and $\Omega_Q\bar{\Omega}_Q$ ($Q=c, b$), can exist as bound states. Among the heavy dibaryon systems, only the $\Xi_Q\Xi_Q$ system with isospin $I=0$ and the $\Sigma_Q\Sigma_Q$ systems with isospin $I=0$ and $I=1$ can exist as bound states. Additionally, we found that the $\Sigma_Q\bar{\Sigma}_Q$ system with $I=2$ and the $\Sigma_Q\Sigma_Q$ system with $I=1$ cannot exist as very deeply bound states. Furthermore, the large mass of heavy baryons reduces the kinetic energy of the system, making it easier to form bound states. Therefore, as shown in Figs. \ref{Charmed baryonium}, \ref{Charmed dibaryon}, \ref{Bottom baryonium}, and \ref{Bottom dibaryon}, the parameter $\alpha$ required to form bound states in the bottom region is smaller than that in the charm region, implying that the binding in the bottom region is deeper than that in the charm region.

However, there is considerable debate among different models regarding whether heavy baryonium and heavy dibaryon systems can exist as bound states, especially for the heavy dibaryon systems. In our model, the contribution of the $\omega$ meson in the $\Lambda_c\Lambda_c$ system is repulsive, and the attractive contribution of the $\sigma$ meson is insufficient to form a bound state in the $\Lambda_c\Lambda_c$ system. Nevertheless, in many other models \cite{Lu:2017dvm, Chen:2017vai, Gerasyuta:2011zx}, the $\Lambda_c\Lambda_c$ system can exist as a bound state. In Refs. \cite{Meguro:2011nr, Huang:2013rla, Oka:2013xxa}, the $\Lambda_c\Lambda_c$ system cannot be a bound state by itself but it is shown that the coupling to the strongly attractive $\Sigma_c^{(\ast)}\Sigma_c^{(\ast)}$ system may lead to a state below the $\Lambda_c\Lambda_c$ threshold. In the one-boson-exchange model \cite{Lee:2011rka}, the $\Xi'_c\Xi'_c$ system and the $\Omega_c\Omega_c$ system can exist as shallow bound states, while the $I=2$ $\Sigma_c\bar{\Sigma}_c$ system can exist as a deeply bound state. Therefore, the existence of these bound states requires further theoretical studies and experimental verification.

The charmed baryonium bound states can be studied via $B$ decays and $e^+e^-$ collisions at LHCb, RHIC, Belle II, and BES$\mathrm{\uppercase\expandafter{\romannumeral3}}$. With the upcoming BEPCII upgrade to 5.6 GeV by the end of 2024, as well as the completion of PANDA and the Super Tau-Charm Factory, the detailed study of charmed baryon-antibaryon bound states will become possible. Compared with the production of charmed baryon-antibaryon bound states, the production of charmed dibaryon bound states is significantly more challenging and faces immense difficulties, although it can still occur at LHC and RHIC. Charmed dibaryon bound states are highly stable because their constituent particles primarily decay through weak interactions, leading to long lifetimes (except for $\Sigma_c$ which primarily decays via the strong process $\Sigma_c \rightarrow \Lambda_c \pi$). However, due to the larger masses, bound states in the bottom region are more difficult to be produced than those in the charm region.

\acknowledgments
This work was supported by National Natural Science Foundation of China (Project Nos. 12105149, 12405115 12475096 and 12275024).

\begin{appendix}

\section{The flavour wave functions}

For the isospin conventions, we use the following ones:
\begin{equation}
    \vert u\rangle=\Bigg{\vert} \frac12,\frac12\Bigg{\rangle},\quad\vert d\rangle=\Bigg{\vert} \frac12,-\frac12\Bigg{\rangle},\quad \vert \bar{u}\rangle=\Bigg{\vert} \frac12,-\frac12\Bigg{\rangle},\quad \vert \bar{d}\rangle=-\Bigg{\vert} \frac12,\frac12\Bigg{\rangle}.
\end{equation}
Then, we have 
\begin{equation}
    \begin{split}
        \big{\vert} \Lambda_c^+\big{\rangle}&=\vert0,0\rangle,\quad\big{\vert} \Lambda_c^-\big{\rangle}=-\vert0,0\rangle,\quad \big{\vert} \Sigma_c^{++}\big{\rangle}=\vert1,1\rangle,\quad\big{\vert} \Sigma_c^{--}\big{\rangle}=\vert1,-1\rangle,\\
        \big{\vert} \Sigma_c^{+}\big{\rangle}&=\vert1,0\rangle,\quad\big{\vert} \Sigma_c^{-}\big{\rangle}=-\vert1,0\rangle,\quad\big{\vert} \Sigma_c^{0}\big{\rangle}=\vert1,-1\rangle,\quad\big{\vert} \bar{\Sigma}_c^{0}\big{\rangle}=\vert1,1\rangle,\\
        \big{\vert} \Xi_c^{(')+}\big{\rangle}&=\vert\frac12,\frac12\rangle,\quad\big{\vert} \Xi_c^{(')-}\big{\rangle}=\vert\frac12,-\frac12\rangle,\quad\big{\vert} \Xi_c^{(')0}\big{\rangle}=\vert\frac12,-\frac12\rangle,\quad\big{\vert} \bar{\Xi}_c^{(')0}\big{\rangle}=-\vert\frac12,\frac12\rangle,\\
        \big{\vert} \Omega_c^{0}\big{\rangle}&=\vert0,0\rangle,\quad\big{\vert} \bar{\Omega}_c^{0}\big{\rangle}=\vert0,0\rangle.
    \end{split}
\end{equation}
The flavour wave functions of the charmed baryon and anti-charmed baryon (charmed baryon) systems can be construct with the Clebsch-Gordan coefficients and above conventions,
\begin{equation}
    \begin{split}
        \Psi_{\Lambda_c\bar{\Lambda}_c}^{0,0}&=-\big{\vert}\Lambda_c\bar{\Lambda}_c\big{\rangle},\\      
        \Psi_{\Sigma_c\bar{\Sigma}_c}^{0,0}&=\frac{1}{\sqrt{3}}\big{\vert}\Sigma_c^{++}\Sigma_c^{--}+\Sigma_c^{+}{\Sigma}_c^{-}+\Sigma_c^{0}\bar{\Sigma}_c^{0}\big{\rangle},\\
        \Psi_{\Sigma_c\bar{\Sigma}_c}^{1,0}&=\frac{1}{\sqrt{2}}\big{\vert}\Sigma_c^{++}{\Sigma}_c^{--}-\Sigma_c^{0}\bar{\Sigma}_c^{0}\big{\rangle},\
        \Psi_{\Sigma_c\bar{\Sigma}_c}^{1,1}=-\frac{1}{\sqrt{2}}\big{\vert}\Sigma_c^{++}{\Sigma}_c^{-}+\Sigma_c^{+}\bar{\Sigma}_c^{0}\big{\rangle},\
        \Psi_{\Sigma_c\bar{\Sigma}_c}^{1,-1}=\frac{1}{\sqrt{2}}\big{\vert}\Sigma_c^{+}{\Sigma}_c^{--}+\Sigma_c^{0}{\Sigma}_c^{-}\big{\rangle},\\
        \Psi_{\Sigma_c\bar{\Sigma}_c}^{2,0}&=\frac{1}{\sqrt{6}}\big{\vert}\Sigma_c^{++}{\Sigma}_c^{--}-2\Sigma_c^{+}{\Sigma}_c^{-}+\Sigma_c^{0}\bar{\Sigma}_c^{0}\big{\rangle},\
        \Psi_{\Sigma_c\bar{\Sigma}_c}^{2,1}=-\frac{1}{\sqrt{2}}\big{\vert}\Sigma_c^{++}{\Sigma}_c^{-}-\Sigma_c^{+}\bar{\Sigma}_c^{0}\big{\rangle},\\        \Psi_{\Sigma_c\bar{\Sigma}_c}^{2,2}&=\big{\vert}\Sigma_c^{++}\bar{\Sigma}_c^{0}\big{\rangle},\ \Psi_{\Sigma_c\bar{\Sigma}_c}^{2,-1}=\frac{1}{\sqrt{2}}\big{\vert}\Sigma_c^{+}\Sigma_c^{--}-{\Sigma}_c^{0}\Sigma_c^{-}\big{\rangle},\
        \Psi_{\Sigma_c\bar{\Sigma}_c}^{2,-2}=\big{\vert}{\Sigma}_c^{0}\Sigma_c^{--}\big{\rangle},\\
         \Psi_{\Xi_c^{(')}\bar{\Xi}_c^{(')}}^{0,0}&=\frac{1}{\sqrt{2}}\big{\vert}\Xi_c^{(')+}\bar{\Xi}_c^{(')-}+\Xi_c^{(')0}\bar{\Xi}_c^{(')0 }\big{\rangle},\\
         \Psi_{\Xi_c^{(')}\bar{\Xi}_c^{(')}}^{1,0}&=\frac{1}{\sqrt{2}}\big{\vert}\Xi_c^{(')+}{\Xi}_c^{(')-}-\Xi_c^{(')0}\bar{\Xi}_c^{(')0}\big{\rangle},\ \Psi_{\Xi_c^{(')}\bar{\Xi}_c^{(')}}^{1,1}=-\big{\vert}\Xi_c^{(')+}\bar{\Xi}_c^{(')0}\big{\rangle},\ \Psi_{\Xi_c^{(')}\bar{\Xi}_c^{(')}}^{1,-1}=\big{\vert}\Xi_c^{(')0}{\Xi}_c^{(')-}\big{\rangle},\\
       \Psi_{\Omega_c\bar{\Omega}_c}^{0,0}&=\big{\vert}\Omega_c^0\bar{\Omega}_c^0\big{\rangle},\\
    \end{split}
\end{equation}

and
\begin{equation}
    \begin{split}
      \Psi_{\Lambda_c\Lambda_c}^{0,0}&=\big{\vert}\Lambda_c{\Lambda}_c\big{\rangle},\\      
        \Psi_{\Sigma_c{\Sigma}_c}^{0,0}&=\frac{1}{\sqrt{3}}\big{\vert}\Sigma_c^{++}\Sigma_c^{0}-\Sigma_c^{+}{\Sigma}_c^{+}+\Sigma_c^{0}{\Sigma}_c^{++}\big{\rangle},\\
        \Psi_{\Sigma_c{\Sigma}_c}^{1,0}&=\frac{1}{\sqrt{2}}\big{\vert}\Sigma_c^{++}{\Sigma}_c^{0}-\Sigma_c^{0}{\Sigma}_c^{++}\big{\rangle},\
        \Psi_{\Sigma_c{\Sigma}_c}^{1,1}=\frac{1}{\sqrt{2}}\big{\vert}\Sigma_c^{++}{\Sigma}_c^{+}-\Sigma_c^{+}{\Sigma}_c^{++}\big{\rangle},\
        \Psi_{\Sigma_c{\Sigma}_c}^{1,-1}=\frac{1}{\sqrt{2}}\big{\vert}\Sigma_c^{+}{\Sigma}_c^{0}-\Sigma_c^{0}{\Sigma}_c^{+}\big{\rangle},\\
        \Psi_{\Sigma_c{\Sigma}_c}^{2,0}&=\frac{1}{\sqrt{6}}\big{\vert}\Sigma_c^{++}{\Sigma}_c^{0}+2\Sigma_c^{+}{\Sigma}_c^{+}+\Sigma_c^{0}{\Sigma}_c^{++}\big{\rangle},\
        \Psi_{\Sigma_c{\Sigma}_c}^{2,1}=\frac{1}{\sqrt{2}}\big{\vert}\Sigma_c^{++}{\Sigma}_c^{+}+\Sigma_c^{+}{\Sigma}_c^{++}\big{\rangle},\\        \Psi_{\Sigma_c{\Sigma}_c}^{2,2}&=\big{\vert}\Sigma_c^{++}{\Sigma}_c^{++}\big{\rangle},\ \Psi_{\Sigma_c{\Sigma}_c}^{2,-1}=\frac{1}{\sqrt{2}}\big{\vert}\Sigma_c^{+}\Sigma_c^{0}+{\Sigma}_c^{0}\Sigma_c^{+}\big{\rangle},\
        \Psi_{\Sigma_c{\Sigma}_c}^{2,-2}=\big{\vert}{\Sigma}_c^{0}\Sigma_c^{0}\big{\rangle},\\
         \Psi_{\Xi_c^{(')}{\Xi}_c^{(')}}^{0,0}&=\frac{1}{\sqrt{2}}\big{\vert}\Xi_c^{(')+}{\Xi}_c^{(')0}-\Xi_c^{(')0}{\Xi}_c^{(')+ }\big{\rangle},\\
         \Psi_{\Xi_c^{(')}{\Xi}_c^{(')}}^{1,0}&=\frac{1}{\sqrt{2}}\big{\vert}\Xi_c^{(')+}{\Xi}_c^{(')0}+\Xi_c^{(')0}{\Xi}_c^{(')+}\big{\rangle},\ \Psi_{\Xi_c^{(')}{\Xi}_c^{(')}}^{1,1}=\big{\vert}\Xi_c^{(')+}{\Xi}_c^{(')+}\big{\rangle},\ \Psi_{\Xi_c^{(')}{\Xi}_c^{(')}}^{1,-1}=\big{\vert}\Xi_c^{(')0}{\Xi}_c^{(')0}\big{\rangle},\\
       \Psi_{\Omega_c{\Omega}_c}^{0,0}&=\big{\vert}\Omega_c^0{\Omega}_c^0\big{\rangle},\\
    \end{split}
\end{equation}
The wave functions of the bottom baryonium and bottom dibaryon systems can be obtained analogously.

\end{appendix}

\end{document}